\documentclass[letter,longauth]{aa} 

\usepackage{graphicx}
\usepackage{txfonts}
\usepackage[]{hyperref}
\usepackage{graphicx}   
\usepackage{amsmath}    
\usepackage{amssymb}    
\usepackage{multicol}        
\usepackage{bm}         
\usepackage{pdflscape}  
\usepackage{threeparttable}
\newcommand{\pname}{$\pi$~Men\,c}
\newcommand{\sname}{$\pi$~Men}

\newcommand\kepler{\emph{{\it Kepler}}}
\newcommand\gaia{\emph{{\it Gaia}}}
\newcommand\jwst{\emph{{\it JWST}}}
\newcommand\elt{\emph{{\it ELT}}}
\newcommand\tmt{\emph{{\it TMT}}}
\newcommand\gmt{\emph{{\it GMT}}}

\newcommand\tess{\emph{\it TESS}}

\newcommand{\ms}{$\mathrm{m\,s^{-1}}$}
\newcommand{\kms}{$\mathrm{km\,s^{-1}}$}

\newcommand\vsini{$v$\,sin\,$i_\star$}

\newcommand\vmic{$v_{\rm mic}$}
\newcommand\vmac{$v_{\rm mac}$}

\newcommand\teff{$T_{\rm eff}$}

\newcommand\logg{log\,{\it g$_\star$}}

\newcommand{\smass}[1][$M_{\odot}$]{ $ 1.02 \pm 0.03 $~#1} 
\newcommand{\sradius}[1][$R_{\odot}$]{ $1.10 \pm 0.01 $~#1}
\newcommand{\stemp}[1][$\mathrm{K}$]{ $ 5870 \pm 50$~#1 }

\newcommand{\Tzerob}[1][days]{$8325.503055 \pm 0.00077 $~#1} 
\newcommand{\Pb}[1][days]{$6.26834 \pm 0.00024 $~#1}

\newcommand{\bb}[1][ ]{$0.614 \pm 0.0180 $~#1} 
\newcommand{\arb}[1][ ]{$13.10\pm 0.18 $~#1} 
\newcommand{\rrb}[1][ ]{$0.01721 \pm 0.00024 $~#1} 
\newcommand{\kb}[1][${\rm m\,s^{-1}}$]{$1.55 \pm 0.27 $~#1} 
\newcommand{\mpb}[1][$M_{\oplus}$]{$4.52 \pm 0.81 $~#1} 
\newcommand{\rpb}[1][$R_{\oplus}$]{$2.06 \pm 0.03 $~#1} 
\newcommand{\ib}[1][deg]{$87.31 \pm 0.11 $~#1} 
\newcommand{\ab}[1][AU]{$0.06702 \pm 0.00109 $~#1} 
 
\newcommand{\denpb}[1][${\rm g\,cm^{-3}}$]{$2.82 \pm 0.53 $~#1} 
\newcommand{\Teqb}[1][K]{$1147 \pm 12 $~#1}
\newcommand{\ttotb}[1][hours]{$2.969_{-0.032}^{+0.030}$~#1}

\newcommand{\Tzeroc}[1][days]{$6548.2 \pm 2.7 $~#1} 
\newcommand{\Tperic}[1][days]{$6306.1 \pm 4.6 $~#1} 
\newcommand{\Pc}[1][days]{$2091.2 \pm 2.0 $~#1} 
\newcommand{\esinc}[1][ ]{$-0.3919 \pm 0.0076 $~#1} 
\newcommand{\ecosc}[1][ ]{$0.6970 \pm 0.0052 $~#1} 
\newcommand{\kc}[1][${\rm m\,s^{-1}}$]{$195.8 \pm 1.4 $~#1} 
\newcommand{\mpc}[1][$M_{\rm J}$]{$ 9.66 \pm 0.20 $~#1} 
\newcommand{\ec}[1][ ]{$0.6394 \pm 0.0025 $~#1} 
\newcommand{\wc}[1][deg]{$330.65 \pm 0.65 $~#1} 
\newcommand{\ac}[1][AU]{$3.22 \pm 0.03 $~#1} 

\newcommand{\qone}[1][]{$0.33 \pm 0.10 $~#1} 
\newcommand{\qtwo}[1][]{$0.23 \pm 0.10 $~#1}

\newcommand{\ATT}[1][${\rm m\,s^{-1}}$]{$0.0021 \pm 0.0011 $~#1} 
\newcommand{\HSone}[1][${\rm m\,s^{-1}}$]{$10.70916 \pm 0.00039 $~#1} 
\newcommand{\HStwo}[1][${\rm m\,s^{-1}}$]{$10.73156 \pm 0.00071 $~#1} 
\newcommand{\jATT}[1][${\rm m\,s^{-1}}$]{$4.26_{-0.96}^{+1.10} $~#1} 
\newcommand{\jHSone}[1][${\rm m\,s^{-1}}$]{$2.35_{-0.17}^{+0.19} $~#1} 
\newcommand{\jHStwo}[1][${\rm m\,s^{-1}}$]{$1.69_{-0.29}^{+0.39} $~#1} 

\begin{document}

   \title{TESS's first planet:}
   \subtitle{a super-Earth transiting the naked-eye star $\pi$~Mensae}

   \author{D.~Gandolfi\inst{\ref{Torino}},
   O.~Barrag\'an\inst{\ref{Torino}}, 
   J.~H.~Livingston\inst{\ref{tokyo}}, 
   M.~Fridlund\inst{\ref{chalmers},\ref{Leiden}},
   A.~B.~Justesen\inst{\ref{SAC}},
   S.~Redfield\inst{\ref{wesleyan}},
   L.~Fossati\inst{\ref{graz}},
   S.~Mathur\inst{\ref{lalaguna},\ref{IAC}},
   S.~Grziwa\inst{\ref{Koln}},
   J.~Cabrera\inst{\ref{DLR}}, 
   R.~A.~Garc\'ia\inst{\ref{paris1},\ref{paris2}},
   C.~M.~Persson\inst{\ref{chalmers}},
   V.~Van~Eylen\inst{\ref{ivy}},
   A.~P.~Hatzes\inst{\ref{TLS}}, 
   D.~Hidalgo\inst{\ref{lalaguna},\ref{IAC}},
   S.~Albrecht\inst{\ref{SAC}}, 
   L.~Bugnet\inst{\ref{paris1},\ref{paris2}},
   W.~D.~Cochran\inst{\ref{austin}}, 
   Sz.~Csizmadia\inst{\ref{DLR}},
   H.~Deeg.\inst{\ref{lalaguna},\ref{IAC}}, 
   Ph.~Eigmuller\inst{\ref{DLR}}, 
   M.~Endl\inst{\ref{austin}}, 
   A.~Erikson\inst{\ref{DLR}},
   M.~Esposito\inst{\ref{TLS}},
   E.~Guenther\inst{\ref{TLS}}, 
   J.~Korth\inst{\ref{Koln}}, 
   R.~Luque\inst{\ref{lalaguna},\ref{IAC}},
   P.~Monta\~nes~Rodr\'iguez\inst{\ref{lalaguna},\ref{IAC}},
   D.~Nespral\inst{\ref{lalaguna},\ref{IAC}},
   G.~Nowak\inst{\ref{lalaguna},\ref{IAC}},  
   M.~Patzold\inst{\ref{Koln}}, 
   J.~Prieto-Arranz\inst{\ref{lalaguna},\ref{IAC}}
    }

\institute{
        Dipartimento di Fisica, Universit\`a degli Studi di Torino, via Pietro Giuria 1, I-10125, Torino, Italy\label{Torino} \\ \email{davide.gandolfi@unito.it}
         \and Department of Astronomy, Graduate School of Science, The University of Tokyo, Hongo 7-3-1, Bunkyo-ku, Tokyo, 113-0033, Japan \label{tokyo}
         \and Department of Space, Earth and Environment, Chalmers University of Technology, Onsala Space Observatory, 439 92 Onsala, Sweden \label{chalmers}
         \and Leiden Observatory, University of Leiden, PO Box 9513, 2300 RA, Leiden, The Netherlands\label{Leiden}
         \and Stellar Astrophysics Centre, Deparment of Physics and Astronomy, Aarhus University, Ny Munkegrade 120, DK-8000 Aarhus C, Denmark\label{SAC}
         \and Astronomy Department and Van Vleck Observatory, Wesleyan University, Middletown, CT 06459, USA\label{wesleyan}
         \and Space Research Institute, Austrian Academy ofSciences, Schmiedlstrasse 6, A-8041 Graz, Austria \label{graz}
         \and Departamento de Astrofísica, Universidad de La Laguna, E-38206, Tenerife, Spain \label{lalaguna} 
         \and Instituto de Astrofísica de Canarias, C/ Vía Láctea s/n, E-38205, La Laguna, Tenerife, Spain\label{IAC} 
         \and Rheinisches Institut f\"ur Umweltforschung, Abteilung Planetenforschung an der Universit\"at zu K\"oln, Aachener Strasse 209, 50931 K\"oln, Germany\label{Koln}
         \and Institute of Planetary Research, German Aerospace Center, Rutherfordstrasse 2, 12489 Berlin, Germany\label{DLR}
         \and IRFU, CEA, Universit\'e Paris-Saclay, F-91191 Gif-sur-Yvette, France\label{paris1}
         \and Universit\'e Paris Diderot, AIM, Sorbonne Paris Cit\'e, CEA, CNRS, F-91191 Gif-sur-Yvette, France\label{paris2}
         \and Department of Astrophysical Sciences, Princeton University, 4 Ivy Lane, Princeton, NJ, 08544, USA \label{ivy}
         \and Th\"uringer Landessternwarte Tautenburg, Sternwarte 5, D-07778 Tautenberg, Germany\label{TLS}
         \and Department of Astronomy and McDonald Observatory, University of Texas at Austin, 2515 Speedway, Stop C1400, Austin, TX 78712, USA \label{austin}
}
   
\date{Received September 20, 2018; accepted September 28, 2018}

\titlerunning{The planetary system orbiting $\pi$~Mensae}
\authorrunning{Gandolfi et al.}

  \abstract
   {                 
   We report on the confirmation and mass determination of \sname~c, the first transiting planet discovered by NASA's \tess\ space mission. \sname\ is a naked-eye (V=5.65\,mag), quiet G0\,V star that was previously known to host a sub-stellar companion (\sname~b) on a long-period ($P_\mathrm{orb}$\,=\,2091~days), eccentric ($e$\,=\,0.64) orbit. Using \tess\ time-series photometry, combined with \gaia\ data, published UCLES@AAT Doppler measurements, and archival HARPS@ESO-3.6m radial velocities, we found that \sname\,c is a close-in planet with an orbital period of $P_\mathrm{orb}$\,=\,6.27\,days, a mass of $M_\mathrm{c}$\,=\,\mpb, and a radius of $R_\mathrm{c}$\,=\,\rpb. Based on the planet's orbital period and size, \sname~c is a super-Earth located at, or close to, the radius gap, while its mass and bulk density suggest it may have held on to a significant atmosphere. Because of the brightness of the host star, this system is highly suitable for a wide range of further studies to characterize the planetary atmosphere and dynamical properties. We also performed an asteroseismic analysis of the \tess\ data and detected a hint of power excess consistent with the seismic values expected for this star, although this result depends on the photometric aperture used to extract the light curve. This marginal detection is expected from pre-launch simulations hinting at the asteroseismic potential of the \tess\ mission for longer, multi-sector observations and/or for more evolved bright stars.
   }


   \maketitle
%

\section{Introduction}

Successfully launched on 18 April 2018, NASA's Transiting Exoplanet Survey Satellite \citep[\tess; ][]{Ricker2015} will provide us with a leap forward in understanding the diversity of small planets (R$_\mathrm{p}$\,$<$\,4\,$R_\oplus$). Unlike previous space missions, \tess\ is performing an all-sky transit survey focusing on bright stars (5\,$<$\,$V$\,$<$\,11 mag) so that detailed characterizations of the planets and their atmospheres can be performed. In its two-year prime mission, \tess\ observes first the southern and then the northern ecliptic hemisphere. The survey is broken up into 26 anti-solar sky sectors. \tess\ uses four cameras to observe each sector, resulting in a combined field of view of $24^\circ$\,$\times$\,$96^\circ$. The overlap between sectors towards the ecliptic poles provides greater sensitivity to smaller and longer-period planets in those regions of the celestial sphere. \tess\ records full-frame images of its entire field of view every 30 minutes and observes approximately 200\,000 pre-selected main-sequence stars with a cadence of $\sim$2 minutes. The mission will certainly open a new era in the study of close-in small planets, providing us with cornerstone objects amenable to both mass determination -- via Doppler spectroscopy -- and atmospheric characterization -- via transmission spectroscopy with NASA's James Webb space telescope (\jwst) and the next generation of extremely large ground-based telescopes (\elt, \tmt, and  \gmt).

Following a successful commissioning of 3 months, \tess\ started the science operation on 25 July 2018 by photometrically monitoring its first sector (Sector 1), which is centered at coordinates $\alpha$\,=\,352.68$\degr$, $\delta$\,=\,$-$64.85$\degr$ (J2000). Shortly after $\sim$30~days of (almost) continuous observations in Sector 1, 73 transiting planet candidates were detected in the two-minute cadence light curves by the \tess\ team and made available to the scientific community upon registration to a dedicated web portal hosted at the Massachusetts Institute of Technology (MIT) web page\footnote{Available at {https://tess.mit.edu/alerts/}.}.

In this letter, we present the spectroscopic confirmation of $\pi$~Men\,c, the first transiting planet discovered by the \tess\ space mission. The host star is \object{$\pi$~Mensae} (\object{HD\,39091}; Table~\ref{table:1}), a naked-eye (V=5.65 mag), relatively inactive \citep[log\,$R^\prime_\mathrm{HK}$\,=\,$-4.941$;][]{Gray2006} G0\,V star already known to host a sub-stellar companion ($\pi$~Men\,b) on a $\sim$2100-day eccentric (e\,$\approx$\,0.6) orbit \citep{Jones2002}. $\pi$~Men\,c is a 2.06\,$R_\oplus$ planet with an orbital period of 6.27 days. Using \gaia\ photometry, archival HARPS Doppler data, and published UCLES high-precision radial velocities (RVs) we confirmed the planetary nature of the transiting signal detected by \tess\ and derived the planet's mass. We note that in the final stage of preparing this manuscript, an independent investigation of this system was publicly announced by \citet{huang2018}.

\section{\tess\ photometry}
\label{Sect:Tess_Photometry}

We downloaded the \tess\ Sector 1 light curves from the MIT web site. For the \tess\ object of interest \object{TOI-144} (\object{$\pi$~Men}, HD\,39091, \object{TIC\,261136679}), the light curve is provided by the NASA Ames Science Processing Operations Center (SPOC). The time-series includes 18\,036 short-cadence (T$_\mathrm{exp}$\,=\,2 min) photometric measurements. \tess\ observations started on 25 July 2018 and ended on 22 August 2018. We removed any measurements that have a non-zero ``quality'' flag, that is, those suffering from cosmic rays or instrumental issues. We removed any long-term stellar variability by fitting a cubic spline with a width of 0.75 days. We searched the light curve for transit signals using the Box-least-squares algorithm \citep[BLS;][]{Kovacs2002}. We detected the signal of \sname~c with a signal-to-noise ratio (S/N) of 9.1 and our ephemeris is consistent with that reported by the \tess\ team. We did not find any additional transit signal with S/N\,>\,6. We also performed a periodogram and auto-cross-correlation analysis in an attempt to extract the rotation period of the star from the out-of-transit \tess\ light curve, but we found no significant rotation signal in the light curve.

\begin{table}
\caption{Main identifiers, coordinates, and parallax, optical, and infrared magnitudes of \sname.}
\label{table:1}
\centering
\begin{tabular}{c c c}
\hline\hline
\noalign{\smallskip}
Parameter & Value & Source \\
\noalign{\smallskip}
\hline
\noalign{\smallskip}
HD & 39091 &  \\
TIC ID & 261136679 & TIC  \\
TOI ID & 144       & \tess\ Alerts \\
\textit{Gaia} DR2 ID & 4623036865373793408 & \textit{Gaia} DR2 \\
RA (J2000) & 05 37 09.885 & \gaia\ DR2 \\
RA (J2000) & -80 28 08.831 & \gaia\ DR2 \\
$\pi$ & $54.705 \pm 0.0671$ mas & \gaia\ DR2 \\
$V$ & $5.65 \pm 0.01$ & \citet{Mermilliod1987} \\
$B$ & $6.25 \pm 0.01$ & \citet{Mermilliod1987} \\
$J$ & $4.869 \pm 0.272$ & 2MASS \\
$H$  & $4.424 \pm 0.226$ & 2MASS \\
$K_s$ & $4.241 \pm 0.027$ & 2MASS \\
$G$ & $5.4907 \pm 0.0014$ & \gaia\ DR2 \\
$G_{\rm BP}$ & $5.8385 \pm 0.0041$ &  \gaia\ DR2 \\
$G_{\rm RP}$ & $5.0643 \pm 0.0034$ & \gaia\ DR2 \\
\noalign{\smallskip}
\hline
\end{tabular}
\begin{tablenotes}\footnotesize
 \item \emph{Note} -- Values marked with TIC, \gaia\ DR2, and 2MASS are from \citet{Stassun2018}, \citet{GaiaDR2}, and \citet{Cutri2003}, respectively. 
 \end{tablenotes}
\end{table}

\section{Limits on photometric contamination}
\label{sec:contam}

\begin{figure}
\begin{center}
\includegraphics[clip,trim={1.2cm 0.75cm 1.0cm 0.2cm},width=0.9\columnwidth]{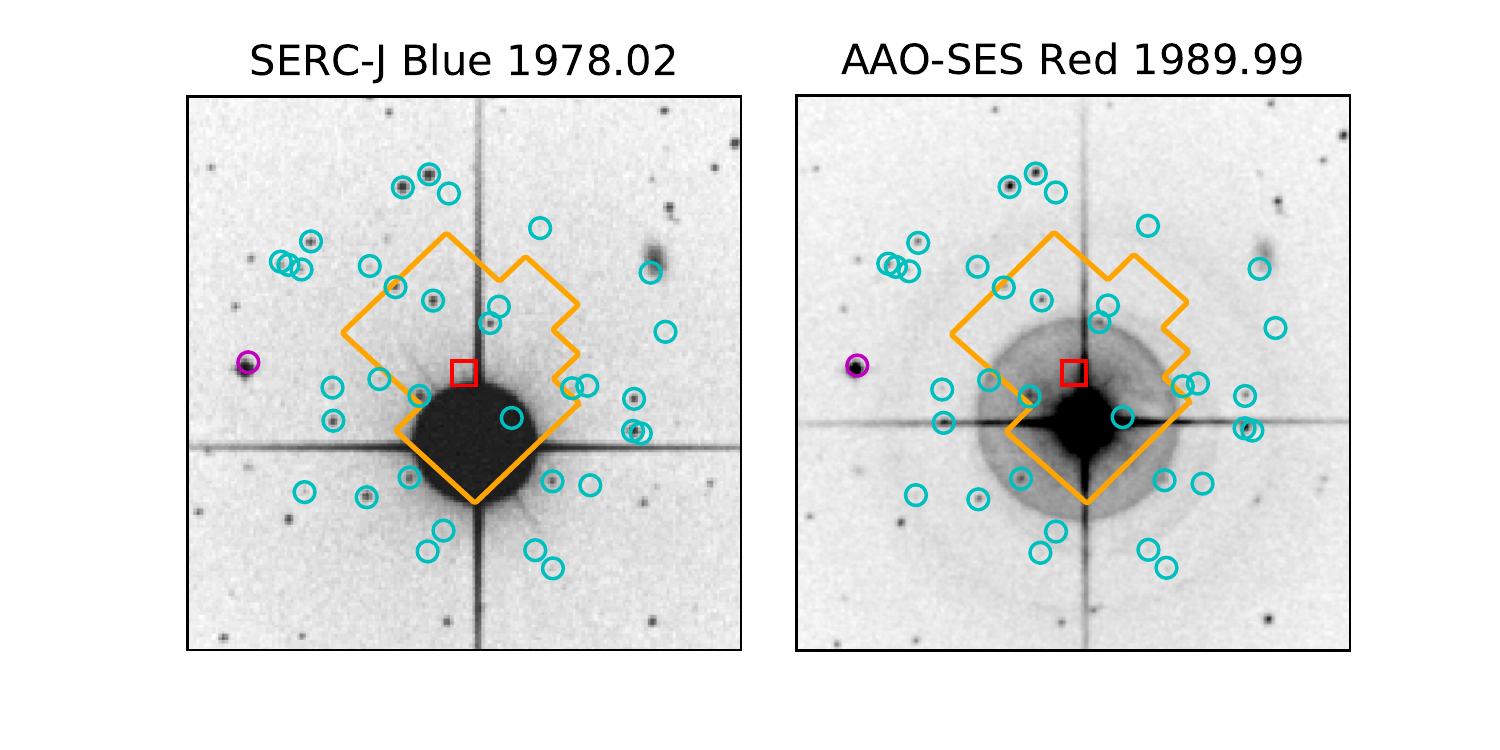}
\caption{5\arcmin$\times$5\arcmin\ archival images with the SPOC photometric aperture overplotted in orange, the \gaia\ DR2 position (J2015.5) of \sname\ indicated by a red square, and other \gaia\ DR2 sources within 2\arcmin\ of \sname\ indicated by circles. The magenta circle indicates the position of \gaia\ DR2 4623036143819289344, a nearby source bright enough to be the host of the observed transit signals (see Section~\ref{sec:contam}), and cyan circles indicate sources that are too faint. \label{fig:dss}}
\end{center}
\end{figure}

As a result of the $\sim$21\arcsec\ pixel scale of the \tess\ detectors, photometric contamination due to chance alignment with a background source is more likely than in previous transit surveys, such as \kepler. We investigated this possibility using archival images of \sname\ from the SERC-J and AAO-SES surveys\footnote{Available at \url{http://archive.stsci.edu/cgi-bin/dss_form}.} and \gaia\ DR2 \citep{GaiaDR2}. The \tess\ photometric aperture used to create the SPOC light curve is approximately 6\,$\times$\,6 \tess\ pixels in extent. We thus executed a query of \gaia\ DR2 centered on the coordinates of \sname\ from the \tess\ Input Catalog\footnote{Available at \url{https://mast.stsci.edu/portal/Mashup/Clients/Mast/Portal.html}.} \citep[TIC;][]{Stassun2018} using a search radius of 2\arcmin. The archival images were taken in 1978 and 1989, and therefore \sname\ appears significantly offset from its current position due to proper motion; no background source is visible near the current position of \sname. Figure~\ref{fig:dss} shows \gaia\ DR2 source positions overplotted on the archival images, along with the SPOC photometric aperture.

Assuming a maximum eclipse depth of 100\%, the measured transit depth (see Section~\ref{Sect.:JointAnalysis}) puts an upper limit on the magnitude of a putative eclipsing binary within the photometric aperture, since a fainter source would be overly diluted  by the flux from \sname. As the \gaia\ $G_\mathrm{RP}$ band-pass is a good approximation to the \tess\ band-pass, we find a limiting magnitude of $G_{\mathrm{RP,max}} = 14.1$ mag. Assuming an aperture radius of 60\arcsec\ (120\arcsec), a simulated stellar population along the line of sight to \sname\ from {\tt TRILEGAL}\footnote{Available at \url{http://stev.oapd.inaf.it/cgi-bin/trilegal}.} \citep{trilegal} implies a frequency of 0.3578 (1.4312) stars brighter than $G_{\mathrm{RP,max}}$. 
Indeed, only one other \gaia\ DR2 source within 2\arcmin\ of \sname\ is brighter than $G_{\mathrm{RP,max}}$, consistent with the expectation from {\tt TRILEGAL}: \gaia\ DR2 4623036143819289344 ($G_\mathrm{RP}$\,=\,12.1644\,$\pm$\,0.0011\,mag, separation\,$\approx$\,118\arcsec).
As this source is clearly outside of the \tess\ photometric aperture, we conclude that \sname\ is the true host of the transit signal as seen by \tess, and that photometric dilution from sources other than \sname\ is negligible.

\section{Custom light curve preparation}
\label{Section:lightcurve}

Having established that \sname\ is the host of the transit signal based on the above analysis of the SPOC aperture and the corresponding light curve, we then performed an analysis of the \tess\ pixel data by downloading the target pixel file (TPF) for \sname\ from the MAST web page\footnote{Available at \url{https://archive.stsci.edu/prepds/tess-data-alerts/}.}. We extracted a series of light curves from the pixel data by producing aperture masks containing different contiguous sets of pixels centered on \sname, which appeared elongated due to CCD blooming preferentially along the columns of the detector. The apertures were produced by computing the 50$^\mathrm{th}$ to 95$^\mathrm{th}$ percentiles of the median image and selecting pixels with median counts above each percentile value. We then selected an optimal aperture by minimizing the 6.5 hour combined differential photometric precision (CDPP) noise metric \citep{2012PASP..124.1279C}. We found that the light curves exhibited elevated noise levels between BJD$-$2450000\,$\approx$\,8347--8350, which corresponded to a deviation in spacecraft pointing. As no transits occur during this time we decided to mask it from each light curve before computing the value of the CDPP noise metric. An aperture somewhat larger than the SPOC aperture yielded the light curve with the best CDPP value, and we use this for our subsequent transit analysis. We note that, although our custom aperture is larger than the SPOC aperture, \sname\ is so much brighter than other nearby stars that photometric dilution remains negligible compared to the uncertainties of the data.

\section{UCLES and HARPS archival spectra}
\label{Section:spectroscopy}

\citet{Jones2002} reported on the detection of a long-period ($P_\mathrm{orb}$\,$\approx$\,2100 days), eccentric ($e$\,$\approx$\,0.6), sub-stellar companion to \sname\ with a minimum mass of $M_\mathrm{b}$\,$\approx$\,10\,$M_\mathrm{Jup}$. Their discovery is based on 28 RV measurements obtained between November 1998 and April 2002 using the UCLES spectrograph mounted at the 3.92-m Anglo-Australian Telescope at Siding Spring Observatory. Fourteen additional UCLES RVs taken between August 2002 and October 2005 were published by \citet{Butler2006}. For the sake of clarity, we list the 42 UCLES RVs in Table~\ref{Table:AAT}.
 
We also retrieved from the ESO public archive 145 high-resolution ($R$\,$\approx$\,$115\,000$) spectra of \sname, taken with the HARPS spectrograph \citep{Mayor2003} mounted at the ESO-3.6m telescope of La Silla observatory (Chile). The observations were carried out between 28 December 2012 and 17 March 2017 UTC, as part of the observing programs 072.C-0488, 183.C-0972, and 192.C-0852. The retrieved data-set includes Echelle and order-merged spectra in flexible image transport system (FITS) format, along with additional FITS files containing the cross-correlation function (CCF) and its bisector, computed from the HARPS pipeline using a G2 numerical mask. From the FITS headers, we extracted the barycentric Julian dates, the RVs, and their uncertainties, along with the full width at half maximum (FWHM) and bisector span (BIS) of the CCF, and the S/N per pixel at 5500\,\AA. On June 2015, the HARPS fiber bundle was upgraded \citep{LoCurto2015}. To account for the RV offset caused by the instrument refurbishment, we treated the HARPS RVs taken before/after June 2015 as two different data sets (Table~\ref{Table:HARPS1} and \ref{Table:HARPS2}). Following \citet{Eastman2010}, we converted the heliocentric Julian dates (HJD\,$_\mathrm{UTC}$) of the UCLES time stamps and the barycentric Julian (BJD\,$_\mathrm{UTC}$) of the HARPS time stamps into barycentric Julian dates in barycentric dynamical time (BJD\,$_\mathrm{TDB}$).

\section{Stellar fundamental parameters}
\label{Sect.:StellarParameters}

We determined the spectroscopic parameters of \sname\ from the co-added HARPS spectrum, which has a S/N per pixel of $\sim$1880 at 5500\,\AA. We used Spectroscopy Made Easy \citep[\texttt{SME};][]{vp96,vf05,Piskunov2017}, a spectral analysis tool that calculates synthetic spectra and fits them to high-resolution observed spectra using a $\chi^2$ minimizing procedure. The analysis was performed with the non-LTE \texttt{SME} version 5.2.2, along with \texttt{ATLAS\,12} one-dimensional model atmospheres \citep{Kurucz2013}. 

In order to estimate the micro-turbulent (\vmic) and macro-turbulent (\vmac) velocities, we used the empirical calibration equations for Sun-like stars from \cite{Bruntt2010b} and \cite{Doyle2014}, respectively. The effective temperature \teff\ was measured fitting the wings of the H$_\alpha$ line \citep{Fuhrmann93,Axer94,Fuhrmann94,Fuhrmann97a,Fuhrmann97b}. We excluded the core of H$_\alpha$ because of its origin in higher layers of stellar photospheres. The surface gravity \logg\ was determined from the wings of the Ca\,{\sc i}~$\lambda$\,6102, $\lambda$\,6122, $\lambda$\,6162\,\AA\ triplet, and the Ca\,{\sc i} $\lambda$\,6439\,\AA\ line. We measured the iron abundance [Fe/H] and projected rotational velocity \vsini\ by simultaneously fitting the  unblended iron lines in the spectral region 5880--6600\,\AA.

We found an effective temperature of \teff\,= $5870\pm50$ K, surface gravity \logg\,=\,$4.33\pm0.09$\,(cgs), and an iron abundance relative to solar of [Fe/H]\,=\,$0.05\pm0.09$~dex. We measured a [Ca/H] abundance of \,$0.07\pm0.10$~dex. The projected rotational velocity was found to be \vsini\,=\,$3.3\pm0.5$~\kms, with \vmic\,=\,$1.06\pm0.10$~\kms\ and \vmac\,=$3.35\pm0.4$~\kms. These values were confirmed by modeling the Na I doublet at 5889.95 and 5895.924~\AA. We detected no interstellar sodium, as expected given the vicinity of the star (d=18.3\,pc).

We used the BAyesian STellar Algorithm \citep[\texttt{BASTA},][]{Aguirre2015} with a large grid of \texttt{GARSTEC} stellar models \citep{Weiss2008} to derive the fundamental parameters of \sname. We built the spectral energy distribution (SED) of the star using the magnitudes listed in Table~\ref{table:1}, and then fitted the SED along with our spectroscopic parameters (\teff, \logg, [Fe/H]) and \gaia\ parallax to a grid of \texttt{GARSTEC} models. Following \citet{Luri2018}, we quadratically added 0.1 mas to the nominal uncertainty of \gaia\ parallax to account for systematic uncertainties of \gaia\ astrometry. We adopted a minimum uncertainty of 0.01 mags for the \gaia\ magnitudes to account for systematic uncertainties in the \gaia\ photometry. Given the proximity of the star (d=18.3\,pc), we assumed no interstellar reddening. 

We found that \sname\ has a mass of $M_\star$\,=\,\smass\ and a radius of $R_\star$\,=\,\sradius, implying a surface gravity of \logg\,=\, 4.36\,$\pm$\,0.02 (cgs), in agreement with the spectroscopically derived value of 4.33\,$\pm$\,0.09. The stellar models constrain the age of the star to be 5.2\,$\pm$\,1.1~Gyr. The fundamental parameters of \sname\ are given in Table~\ref{tab:parameters}. We stress that the uncertainties on the derived parameters are internal to the stellar models used and do not include systematic uncertainties related to input physics or to the bolometric correction.

\section{Seismic analysis}

In order to better characterize the star, we performed an asteroseismic analysis of the SPOC light curve as well as the custom light curve optimized for characterization of the exoplanet (Sect.~\ref{Section:lightcurve}). For the former, the corrections we made consist of three steps. First, we corrected the SPOC flux\footnote{We used the pre-search data conditioning simple aperture photometry \citep{Smith2012}.} performing a robust locally weighted regression as described in \citet{Cleveland1979} in order to smooth long-period variation from the light curve without removing any transit signal. We also calibrated the data following the methods described in \citet{2011MNRAS.414L...6G}. The results of both analyses provided similar seismic results, although the corrections applied were very different. As a second step we removed the transits by folding the light curve at the period of the planet transit and filtering it with a wavelet transform using an ``\`a trous’’ algorithm \citep{book:starck02, book:starck06}. Finally as the last step, the gaps of the resultant light curve were interpolated using inpainting techniques following \citet{2014A&A...568A..10G} and \citet{2015A&A...574A..18P}. For the custom aperture (Sect.~\ref{Section:lightcurve}), the first step consisted of correcting the light curve following only \citet{2011MNRAS.414L...6G} (as the two corrections methods led to similar results) and we applied the same steps two and three.

We used the FliPer metric \citep{2018arXiv180905105B} to estimate \logg\ directly from the global power of the power spectrum density of both light curves. Unfortunately, due to the high level of noise and the filters applied to the light curve to flatten it and properly remove the transits, part of the power below $\sim$100\,$\mu$Hz is removed providing only a loose limit of the value of surface gravity or the frequency of maximum power of the modes. We applied the standard seismic A2Z pipeline \citep{2010A&A...511A..46M} to look for the power excess due to acoustic modes. While the blind search did not yield a significant detection in either light curve, we estimated where we would expect the acoustic modes given the spectroscopic parameters derived in this paper. The modes are expected around 2500~$\mu$Hz. The power spectrum of SPOC light curve shows a slight excess of power around 2600$~\mu$Hz (frequency of maximum power or $\nu_{max}$) and the A2Z pipeline that computes the power spectrum of the power spectrum detects a signal at 119.98\,$\pm$\,9.25\,$\mu$Hz, which could be the large frequency separation ($\Delta\nu$ is the frequency difference between 2 modes of same degree and consecutive orders) with a 95\% confidence level. This value corresponds to the $\Delta\nu$ expected from the global seismic scaling relations for modes at 2607\,$\pm$\,16\,$\mu$Hz. Although quite unlikely, this signal could be due to noise. If we consider that this is a real seismic signature, using $\Delta\nu$, $\nu_{\rm{max}}$, and $T_{\rm{eff}}$, along with the solar scaling relations, we found a stellar mass of $M_\star$\,=\,1.02\,$\pm$\,0.15~$M_\odot$ and a stellar radius of $R_\star$\,=\,1.09\,$\pm$\,0.10~$R_\odot$, in agreement with the spectroscopic values (Sect.\,\ref{Sect.:StellarParameters}). However, the asteroseismic analysis of the custom light curve does not reproduce the previous results, as the seismic signal is not statistically significant. A better knowledge of the optimal aperture mask for asteroseismology is required at this point, which is beyond the scope of this paper. We note that this marginal detection is expected for a star of this magnitude and evolutionary stage observed over the course of one month. Asteroseismic analysis with \tess\ will require multiple-sector data and/or more evolved bright stars, as predicted by pre-launch simulations \citep{2017EPJWC.16001006C}.

\section{Frequency analysis of the Doppler data}
\label{Sect.:FrequencyAnalysis}

We performed a frequency analysis of the UCLES and HARPS RVs in order to search for the presence of the transiting planet in the Doppler data, and look for possible additional periodic signals. The generalized Lomb-Scargle \citep[GLS;][]{Zechmeister2009} periodogram of the combined UCLES and HARPS RVs\footnote{We accounted for the instrumental offsets using the values derived in Sect.\,\ref{Sect.:JointAnalysis}.} shows a very significant peak at the orbital frequency of the outer sub-stellar companion ($f_b$\,=\,0.0005~c/d), with a false-alarm probability (FAP) lower than 10$^{-10}$.

\begin{figure}
\includegraphics[clip,trim={0cm 0cm 0cm 0cm},width=\columnwidth]{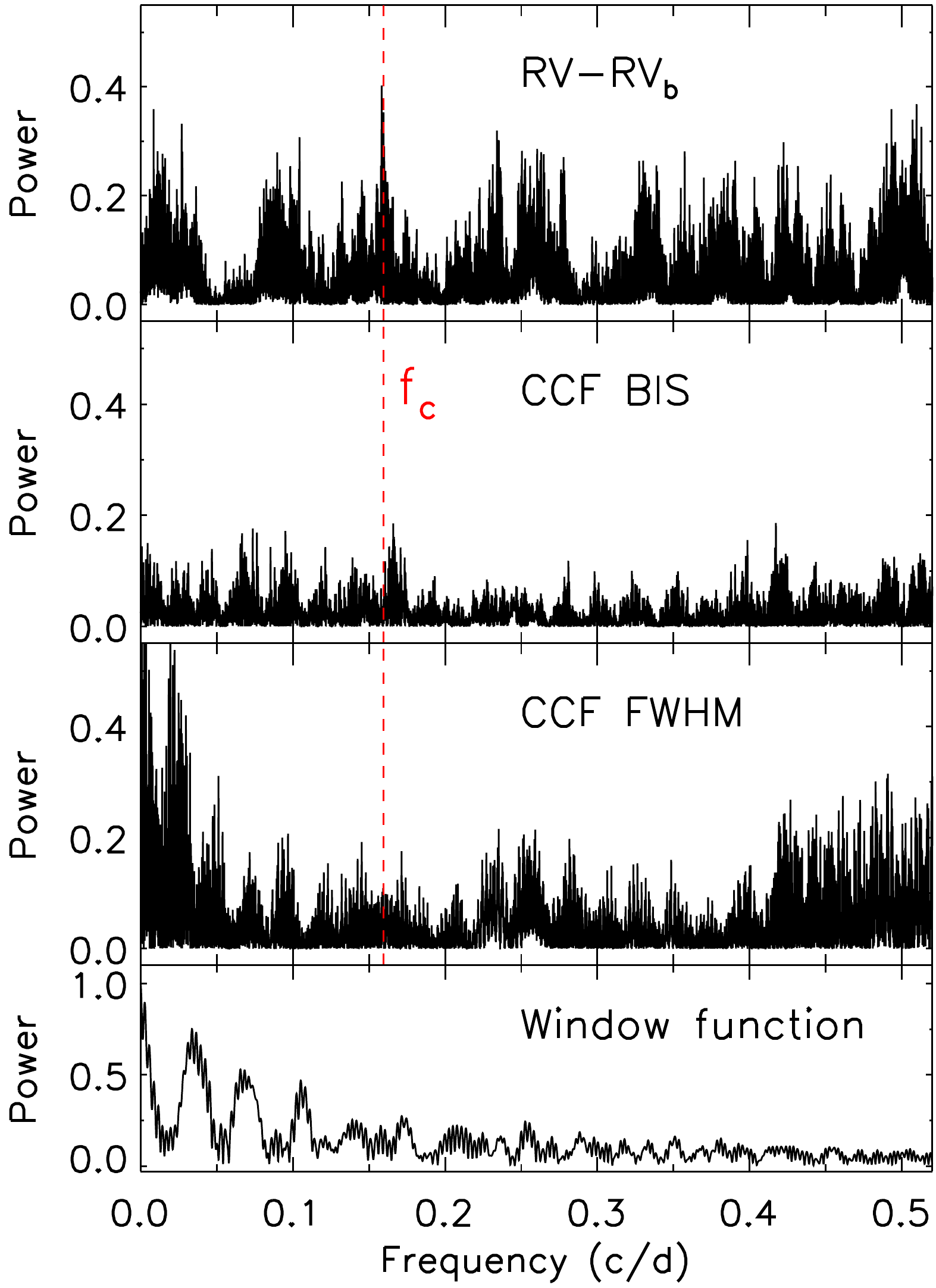}
\caption{\emph{First panel}: GLS periodogram of the UCLAS and HARPS RV residuals following the subtraction of the Doppler reflex motion induced by the outer sub-stellar companion. \emph{Second and third panels}: GLS periodogram of the BIS and FWHM of the HARPS CCF (data acquired with the old fiber bundle). \emph{Fourth panel}: Periodogram of the window function of the combined RV measurements. The dashed vertical red line marks the orbital frequency of the transiting planet ($f_c$\,=\,0.16~c/d). \label{Fig:GLS}}
\end{figure}

The upper panel of Fig.\,~\ref{Fig:GLS} displays the GLS periodogram of the RV residuals following the subtraction of the Doppler signal induced by the outer sub-stellar object. We found that the most significant peak is seen at the frequency of the transit signal detected by \tess\ ($f_c$\,=\,0.16~c/d), with a FAP\,$<$\,$10^{-5}$ and an RV semi-amplitude of $\sim$1.5\,\ms. The peak has no counterparts in the periodograms of the HARPS activity indicators (second and third panels of Fig.\,\ref{Fig:GLS}), suggesting that the signal is induced by the presence of an orbiting planet with a period of 6.27\,days.

We also subtracted the Doppler reflex motion induced by the transiting planet from our RV data and searched the residuals for additional periodic signal but found no peak with FAP\,$<$\,$10^{-4}$.

\section{Joint analysis of the transit and Doppler data}
\label{Sect.:JointAnalysis}

\begin{figure}[!t]
    \begin{center}
    \includegraphics[width=0.48\textwidth]{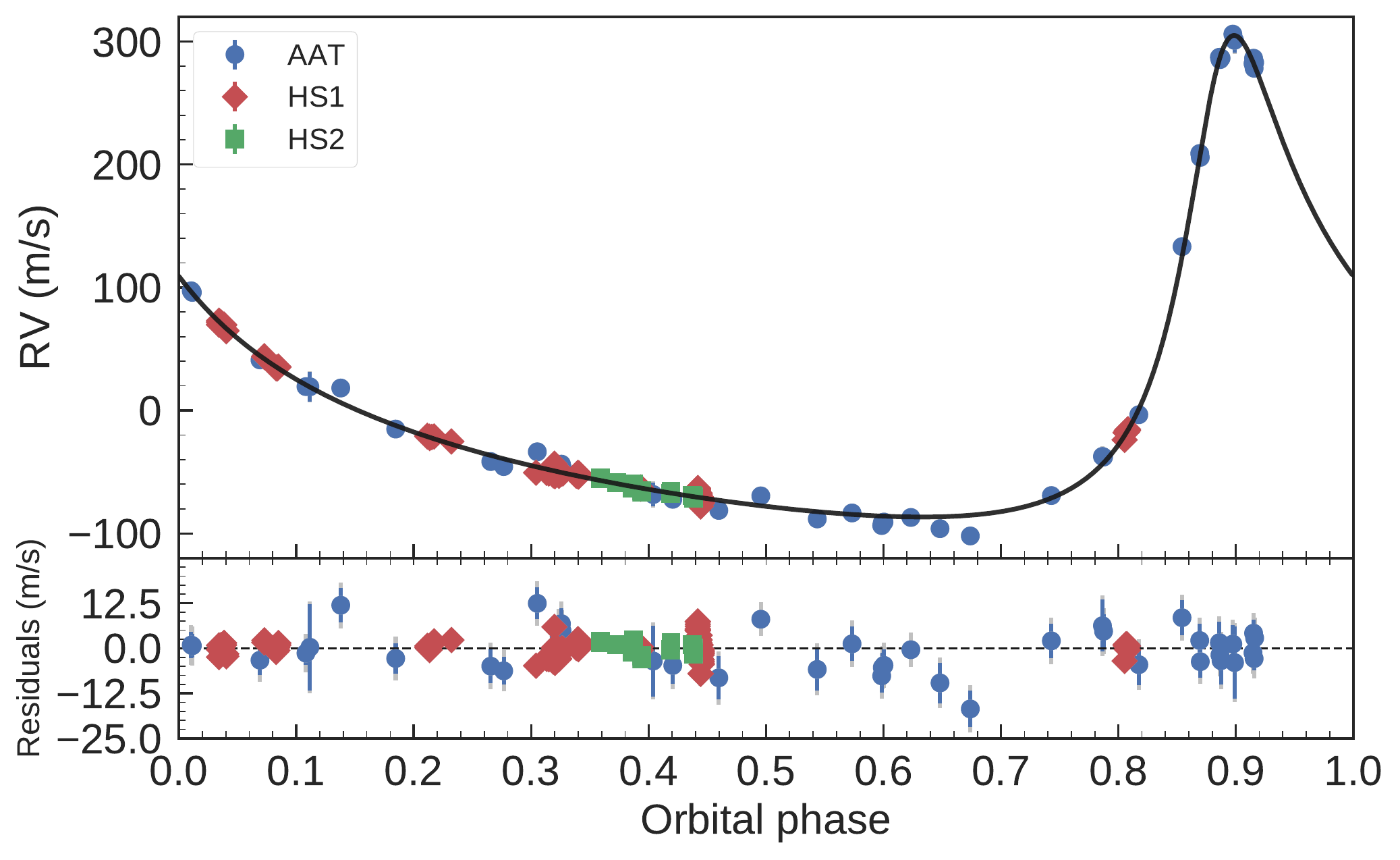}\\
    \includegraphics[width=0.48\textwidth]{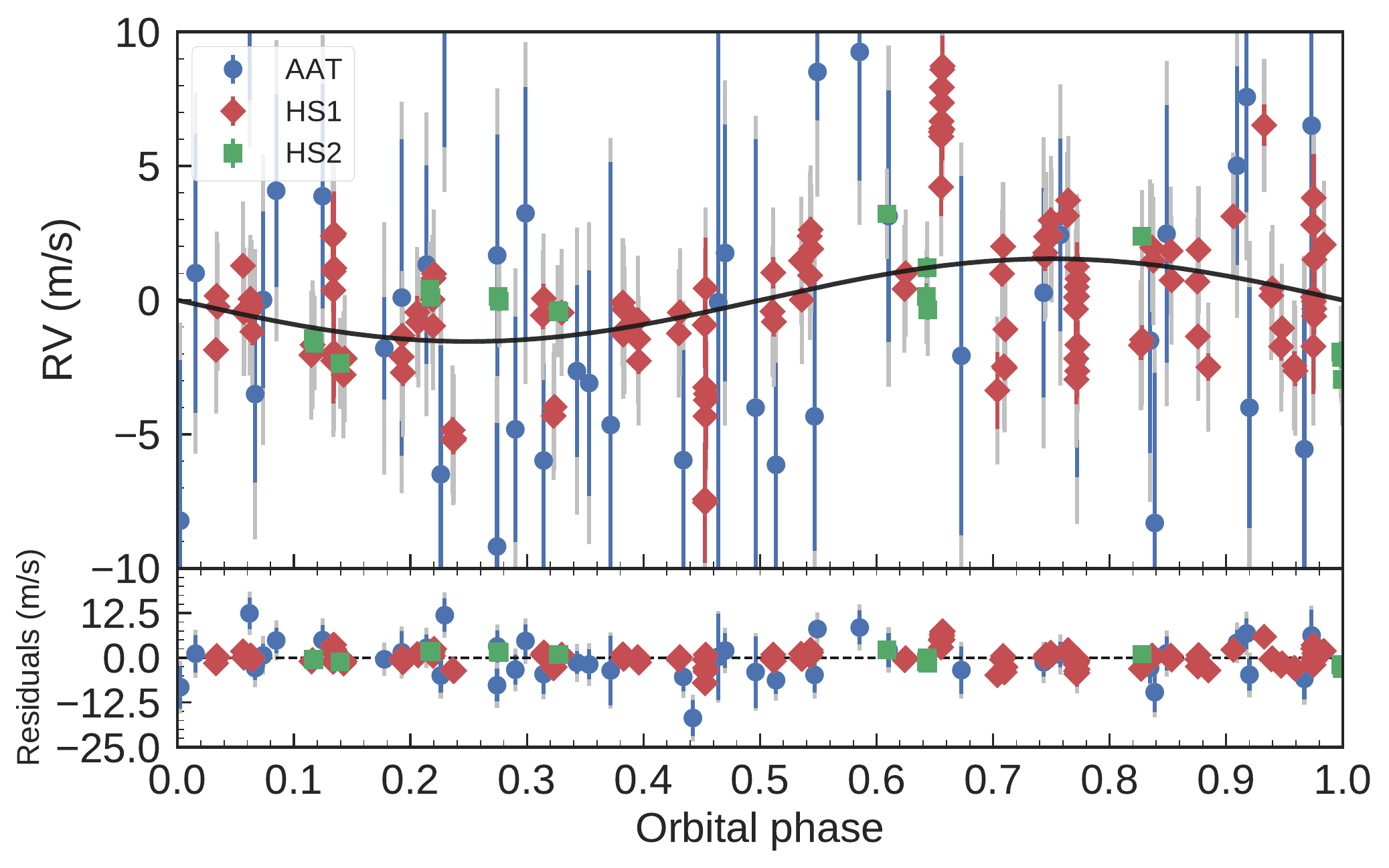}\\
    \includegraphics[clip,trim={0.3cm 0cm -0.45cm 0cm},width=0.483\textwidth]{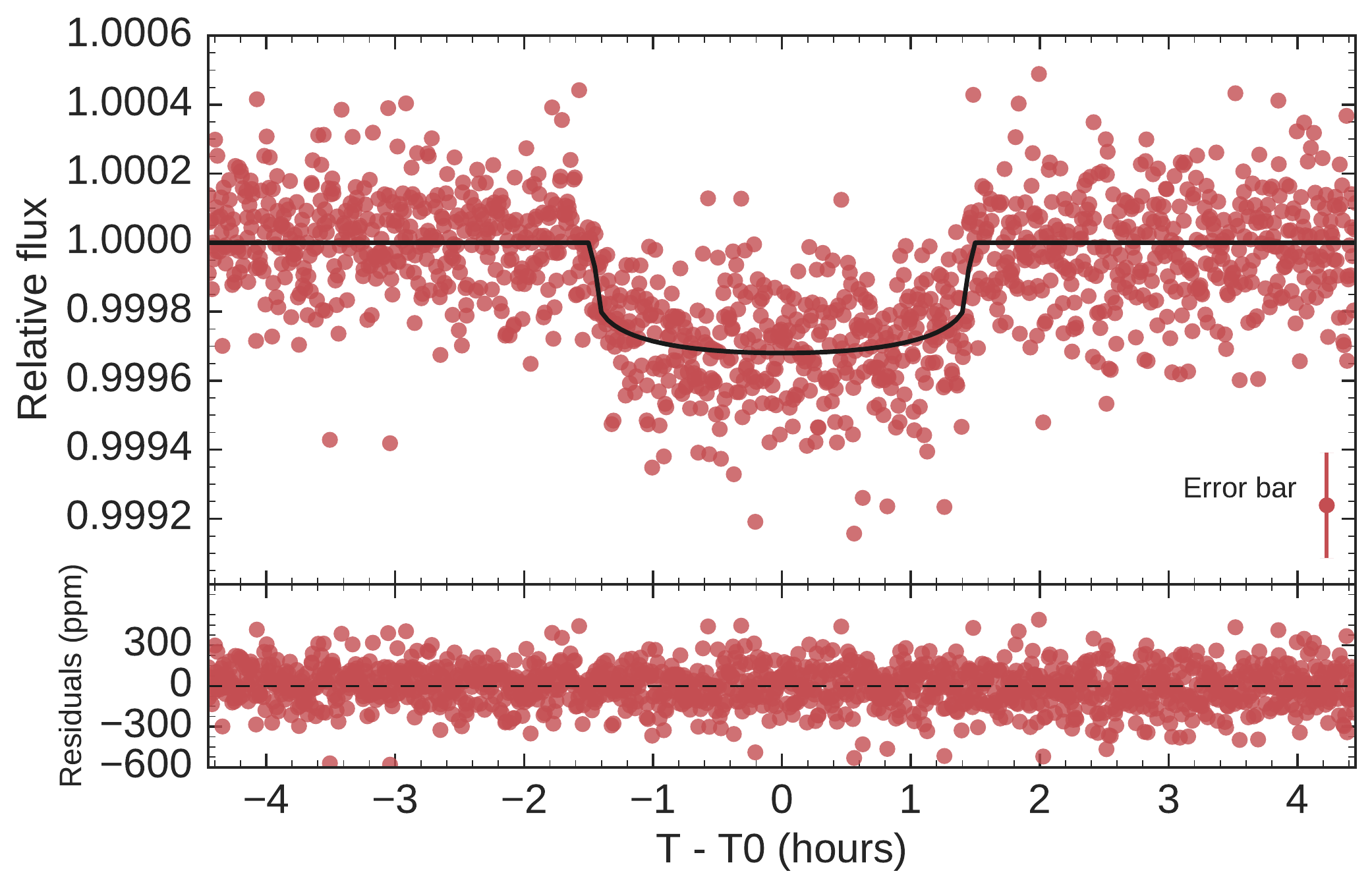}
    \caption{Phase-folded RV curves of \sname~b (upper panel) and c (middle panel), and transit light curve of \sname~c (lower panel). The best-fitting transit and Keplerian models are overplotted with thick black lines. The zero phases of the RV curves of \sname~b (upper panel) and c (lower panel) correspond to the time of inferior conjunction and transit epoch, respectively (Table~\ref{tab:parameters}). The \tess\ data points are shown with red circles (lower panel). The AAT data and the two sets of HARPS RVs (HS1 and HS2) are shown with circles, diamonds, and squares, respectively. The vertical gray lines mark the error bars including jitter. }
    \label{fig:fits}
    \end{center}
\end{figure}

We performed a joint analysis of our custom light curve (Sect.~\ref{Section:lightcurve}) and RV measurements (Sect.~\ref{Section:spectroscopy}) using the software suite \texttt{pyaneti} \citep{Barragan2019}, which allows for parameter estimation from posterior distributions calculated using Markov chain Monte Carlo methods.

We extracted 10 hours of \tess\ data points centered around each of the five transits observed by \tess. The five segments were de-trended using the code \texttt{exotrending} \citep{exotrending}, fitting a second-order polynomial to the out-of-transit data. We used all 187 Doppler measurements presented in Sect.~\ref{Section:spectroscopy} and accounted for the RV offsets between the different instruments and the two HARPS setups.

The RV model consists of two Keplerians to account for the Doppler signal induced by planets $b$ and $c$. We fitted for a RV jitter term for each instrument/setup to account for instrumental noise not included in the nominal uncertainties, and/or to account for any stellar-activity-induced RV variation. We modeled the \tess\ transit light curves using the limb-darkened quadratic model of \citet{2002ApJ...580L.171M}. For the limb darkening coefficients,  we set Gaussian priors using the values derived by \citet{Claret2017} for the \tess\ pass-band. We imposed conservative error bars of 0.1 on both the linear and the quadratic limb-darkening term. A preliminary analysis showed that the transit light curve poorly constrains the scaled semi-major axis ($a/R_\star$). We therefore set a Gaussian prior on $a/R_\star$ using Kepler's third's law, the orbital period, and the derived stellar parameters (Sect.~\ref{Sect.:StellarParameters}). Because the eccentricity of planet $c$ is poorly constrained by the observations, we fixed it to zero for our analysis (see also Section~\ref{sec:Discussion}). We imposed uniform priors for the remaining fitted parameters. Details of the fitted parameters and prior ranges are given in Table~\ref{tab:parameters}. Before performing the final analysis, we ran a numerical experiment to check if the \tess\ two-minute integration time needed to be taken into account following \citet{Kipping2010}. We found no differences in the posterior distributions for fits with and without re-sampling; we therefore proceeded with our analysis without re-sampling. We used 500 independent Markov chains initialized randomly inside the prior ranges. Once all chains converged, we used the last 5\,000 iterations and saved the chain states every ten iterations. This approach generates a posterior distribution of 250\,000 points for each fitted parameter. Table~\ref{tab:parameters} lists the inferred planetary parameters. They are defined as the median and 68\% region of the credible interval of the posterior distributions for each fitted parameter. The transit and RV curves are shown in Fig.~\ref{fig:fits}.

\section{Discussion and conclusion}
\label{sec:Discussion}

\sname\ is a bright (V=5.65\,mag) Sun-like star (SpT=G0\,V) known to host a sub-stellar companion (\sname~b) on a long-period eccentric orbit \citep{Jones2002}. Combining \gaia\ photometry with archival RV measurements we confirmed that the P=6.27\,day transit signal detected in the \tess\ light curve of \sname\ is caused by a \emph{bona fide} transiting super-Earth and derived its mass. \sname~c becomes \tess's first confirmed planet. 

\sname\ joins the growing number of stars known to host both long-period Jupiter analogs and close-in small planets ($R_\mathrm{p}$\,$<$\,4\,$R_\oplus$). \citet{Zhu2018} pointed out that long-period, gas-giant planets are common around stars hosting super-Earths. \citet{Bryan2018} recently found that the occurrence rate of companions in the range $0.5$--$20$\,$M_\mathrm{Jup}$ at 1--20\,AU in systems known to host inner small planets is 39\,$\pm$\,7\%, suggesting that the presence of outer gas giant planets does not prevent the formation of inner Earth- and Neptune-size planets. We performed a dynamical stability analysis of \sname\ using the software \texttt{mercury6} \citep{mercury}. Assuming co-planar orbits, we let the system evolve for 100\,000 yr. For \sname\,b we found negligible changes of the semi-major axis and eccentricity of $< 2.6\times 10^{-3}$\,AU and $3\times 10^{-4}$, respectively. For \pname\ we found no variation larger than $1 \times 10^{-5}$ of its semi-major axis, with changes of its eccentricity $\lesssim$\,0.05.

\begin{figure}
    \centering
    \includegraphics[width=0.49\textwidth]{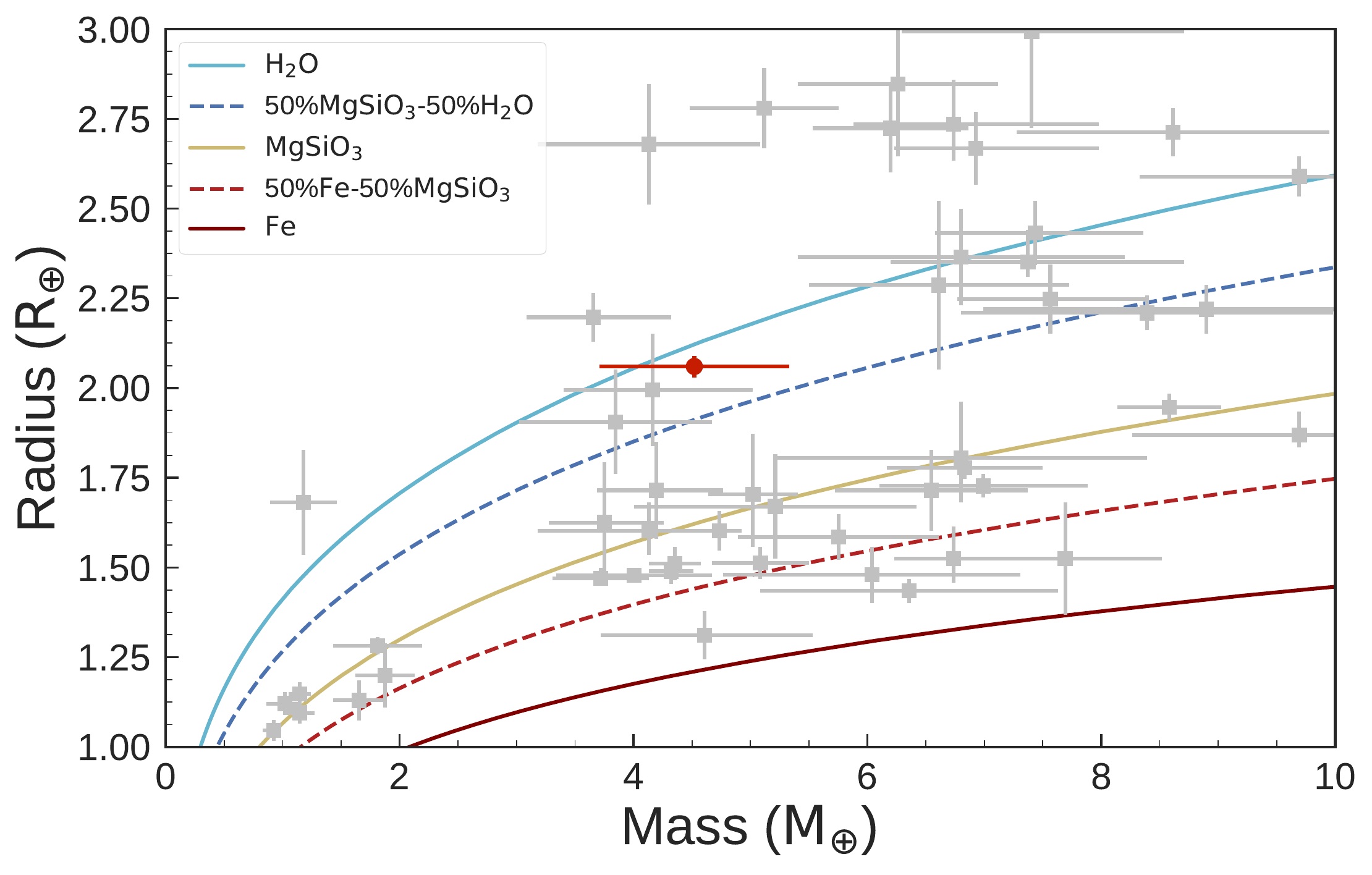}
    \caption{Mass-radius for low-mass ($M_{\rm p}$\,$<$\,10~$M_\oplus$), small ($R_{\rm p}$\,$<$\,3\,$M_\oplus$) planets with mass-radius measurements better than 25\%  \citep[from \url{http://www.astro.keele.ac.uk/jkt/tepcat/};][]{Southworth2007}. Composition models from \citet{Zeng2016} are displayed with different lines and colors. The solid red circle marks the position of \sname~c.
    }
    \label{Fig:MassRadiusDiagram}
\end{figure}

\begin{figure*}
\begin{center}
\includegraphics[clip,trim={2.2cm 12.5cm 2.1cm 3.3cm},width=1.0\columnwidth]{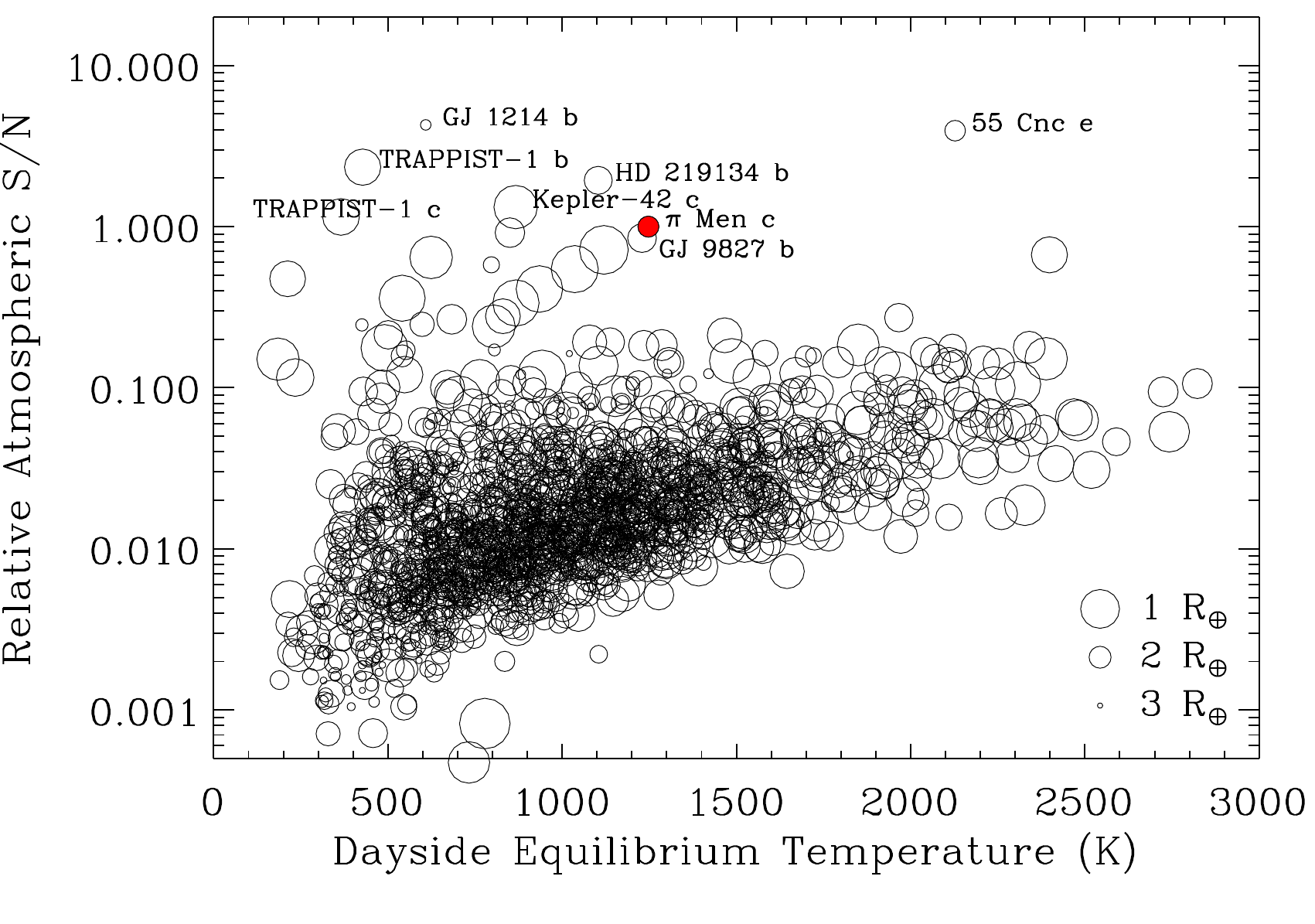}
\includegraphics[clip,trim={2.2cm 12.5cm 2.1cm 3.3cm},width=1.0\columnwidth]{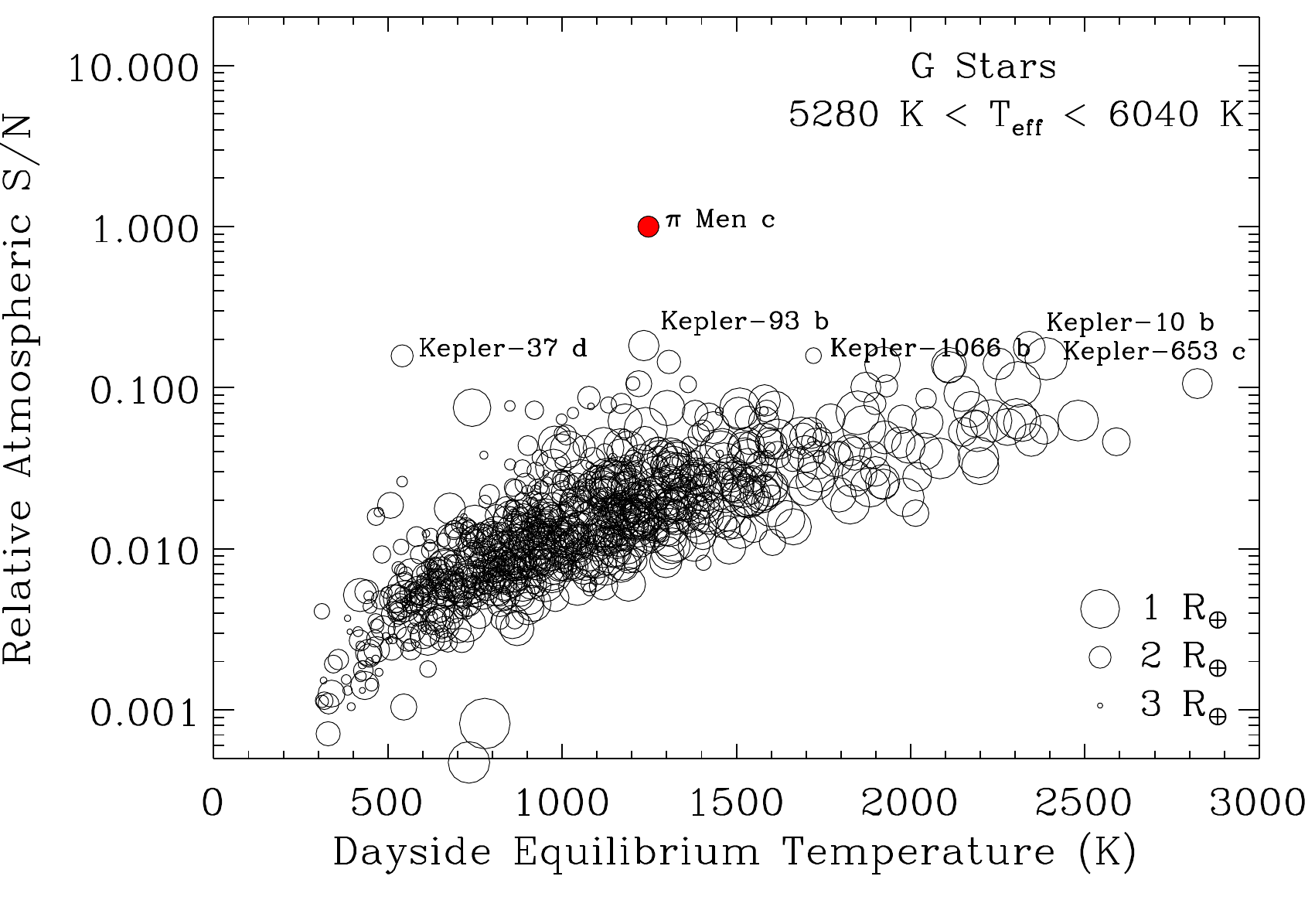}
\caption{\emph{Left Panel}: Relative S/N of an atmospheric signal for all exoplanets with $R_\mathrm{p}$\,$<$\,3~$R_\oplus$ as a function of planetary equilibrium temperature. The symbol size is inversely proportional to the planetary radius, emphasizing those closest to an Earth radius. The \sname~c planet is used as the atmospheric signal reference and it is indicated by the filled circle. It is among the top ten most favorable planets for atmospheric characterization. \emph{Right panel}: As  in the left panel, simply limited to solar type host stars (i.e., G-type stars; 5280\,<\,\teff\,<\,6040 K). The \sname~c planet is by far the most favorable planet around such a star for atmospheric characterization. The other optimal atmospheric targets all transit K and M stars. For this reason, the coronal and stellar wind properties that interact with the \sname~c atmosphere may be distinctly different to those experienced by the rest of the sample. \label{Fig:SNR_Plot_pi_men}}
\end{center}
\end{figure*}

The actual orientation of the outer planet's orbit is unknown. While we know the inner planet's inclination, because it transits, its eccentricity is poorly constrained by the data. Compact multi-planet systems have been observed to have near-zero eccentricities \citep[e.g.,][]{hadden2014,vaneylen2015,xie2016}. However, planets with only a single transiting planet  often appear to be ``dynamically hotter'', and many have a non-zero eccentricity, which can, for example, be described by the positive half of a zero-mean Gaussian distribution, with a dispersion $\sigma_e = 0.32 \pm 0.06$ \citep{vaneylen2018ecc}. The outer planet, $\pi$ Men c, has an orbital eccentricity of $\sim$0.64. A far-out giant planet, such as planet $c$, may in fact increase the orbital eccentricity of a close-in super-Earth, such as planet $b$ \citep[see, e.g.,][]{mustill2017,hansen2017,huang2017}. Following \cite{vaneylen2018ecc}, we found an orbital eccentricity based on the transit data alone of $[0,0.45]$ at 68\% confidence. Because the current RV observations cannot constrain the eccentricity either, we fixed it to zero in the above analysis (see Section~\ref{Sect.:JointAnalysis}). The brightness of the host star makes this planetary system an exciting target for further RV follow-up to measure the inner planet's eccentricity. 

The transiting planet \sname~c has a mass of $M_\mathrm{c}$\,=\,\mpb\ and a radius of $R_\mathrm{c}$\,=\,\rpb, yielding a mean density of $\rho_\mathrm{c}$=\denpb. Figure~\ref{Fig:MassRadiusDiagram} shows the mass-radius diagram for small planets whose masses have been determined with a precision better than 25\,\%. Theoretical models from \citet{Zeng2016} are overplotted using different lines and colors. The position of \sname~c suggests a composition of Mg silicates and water. Alternatively, the planet might have  a solid core surrounded by a gas envelope. At short orbital periods, super-Earths and sub-Neptunes are separated by a radius gap at $\approx$1.8~$R_\oplus$ \citep{fulton2017,VanEylen2018}. The exact location of the radius gap is observed to be a function of the orbital period \citep{VanEylen2018}, as predicted by models of photo-evaporation \citep[e.g.,][]{owen2013,lopez2013}. \cite{VanEylen2018} found that the radius gap is located at $\log R =  m \times \log P + a$, where $m = -0.09^{+0.02}_{-0.04}$ and $a = 0.37^{+0.04}_{-0.02}$. At the orbital period of \sname~c, that is,\ $P_\mathrm{orb} = 6.27$ days, the radius gap is then located at $R_\mathrm{p}$\,=\,1.99\,$\pm$\,0.20~$R_\oplus$. This implies that \sname~c, with a radius of $R_\mathrm{c}$\,=\,\rpb, is located just around the radius gap, although the measured density suggests that the planet may have held on to (part of) its atmosphere.

The naked-eye brightness of \sname\ immediately argues that any transiting planet will be attractive for atmospheric characterization. Observations of a planetary atmosphere through transmission spectroscopy during transit provide opportunities to measure the extent, kinematics, abundances, and structure of the atmosphere \citep{Seager2010}. Such measurements can be utilized to address fundamental questions such as planetary atmospheric escape and interactions with the host star \citep{Cauley2017}, formation and structure of planetary interiors \citep{Owen1999}, planetary and atmospheric evolution \citep{Oberg2011}, and biological processes \citep{Meadows2010}.

The left panel of Fig.\,\ref{Fig:SNR_Plot_pi_men} displays a relative atmospheric detection S/N metric normalized to \sname~c for all known small exoplanets with $R_p$\,<\,3 $R_\oplus$. The sample is taken from the Exoplanet Orbit Database\footnote{Available at \url{exoplanets.org}.} as of September 2018. The atmospheric signal is calculated in a similar way to \citet{Gillon2016} and \citet{Niraula2017}. In particular, the relative atmospheric S/N plotted in Fig.\,\ref{Fig:SNR_Plot_pi_men} is calculated from Eq.\,1a and 1b from \citet{Niraula2017}. This calculation assumes similar atmospheric properties (e.g., Bond albedo, mean molecular weight) for all planets. Large atmospheric signals result from hot, extended atmospheres of planets that transit small, cool stars. For this reason, planets transiting such stars as \object{GJ\,1214} and \object{TRAPPIST-1} are excellent targets for this kind of study. Nevertheless, the brightness of the host star along with the period and duration of the transit also significantly contribute to the ability to build up a sufficiently high S/N to detect atmospheric signals. We used the J-band flux \citep[e.g., H$_2$O measurements with \jwst;][]{Beichman2014} and weighted the metric to optimize the S/N ratio over a period of time rather than per transit.

In the context of all small planets ($R_\mathrm{p}$\,$<$\,3\,$R_\oplus$), \sname~c has the seventh strongest atmospheric signal,  behind \object{GJ\,1214~b}, \object{55\,Cnc~e}, the \object{TRAPPIST-1~b} and \object{c} planets, \object{HD219134~b}, and \object{Kepler-42~c}. However, \sname\ is unique among this notable group of stars in that it is the only G-type star (Fig.~\ref{Fig:SNR_Plot_pi_men}, right panel). All of the other exoplanets transit K- or M-type stars. The brightness of \sname\ is able to overcome the disadvantage of a small planet transiting a slightly larger star, to provide the best opportunity of probing the atmospheric properties of a super-Earth orbiting a solar type star. Given the significant changes in the structure of stellar coronae and stellar winds between G- and M-type stars, the atmospheric properties and evolution of \sname~c may be distinctly different from the atmospheres detected around the sample of very-low-mass M-type stars (e.g., \object{GJ\,1214} and \object{TRAPPIST-1}). For example, the TRAPPIST-1 e, f, and g planets essentially orbit within the stellar corona of the host star and may be subject to a substantial stellar wind, which will result in a strong injection of energy in the atmosphere and may prevent the formation of a significant atmosphere \citep{Cohen2018}. When inferring the properties of coronae and winds of stars other than the Sun, we often have to use poorly constrained models and empirical correlations, the validity of which are best for stars that are quite similar to the Sun. In this respect, \sname\ is a unique laboratory because of its greater similarities to the Sun with respect to all the other stars known to host mini-Neptunes and Super-Earths amenable to multi-wavelength atmospheric characterization. 

We further studied the long-term stability of a possible hydrogen-dominated atmosphere by estimating the mass-loss rates. To this end, we employed the interpolation routine described by \citet{Kubyshkina2018}, which interpolates the mass-loss rate among those obtained with a large grid of one-dimensional upper-atmosphere hydrodynamic models for super-Earths and sub-Neptunes. Employing the values listed in Table~\ref{tab:parameters} and a Sun-like high-energy emission, which is a reasonable assumption given that \sname\ has a temperature and age similar to those of the Sun, we obtained a mass-loss rate of $1.2\times10^{10}$~g\,s$^{-1}$, which corresponds to $\sim$1.4\% of the estimated planetary mass per gigayear. This indicates that the question of whether the planet still holds a hydrogen-dominated atmosphere or not greatly depends on the initial conditions, namely how much hydrogen the planet accreted during its formation. If the planet originally accreted a small hydrogen-dominated atmosphere with a mass of only a few percent of the total planetary mass, we can expect it to be, for the vast majority, lost, particularly taking into account that the star was more active in the past. In contrast, a significant hydrogen mass fraction would still be present if the planet originally accreted a large amount of hydrogen. The inferred bulk density hints at the possible presence of a hydrogen-dominated atmosphere, but it does not give a clear indication. Ultraviolet observations aiming at detecting hydrogen Ly$_\alpha$ absorption and/or carbon and oxygen in the upper planetary atmosphere would be decisive in identifying its true nature.

While this manuscript was in preparation, \citet{huang2018} publicly reported an independent analysis of the \sname\ system, using a custom \tess\ photometry pipeline. They derived a mass and radius for \pname\ consistent within less than 1-$\sigma$ with our measurements.

\begin{acknowledgements}

Davide Gandolfi is lovingly grateful to Conny Konnopke for her unique support during the preparation of this paper, and her valuable suggestions and comments. J.H.L. gratefully acknowledges the support of the Japan Society for the Promotion of Science (JSPS) Research Fellowship for Young Scientists. M.F. and C.M.P. gratefully acknowledge the support of the Swedish National Space Board. A.P.H., Sz.Cs., S.G., J.K., M.P., and M.E. acknowledge support by DFG (Deutsche Forschungsgemeinschaft) grants HA 3279/12-1, PA525/18-1, PA525/19-1, PA525/20-1, and RA 714/14-1 within the DFG Schwerpunkt SPP 1992, ``Exploring the Diversity of Extrasolar Planets.'' We acknowledge the use of \tess\ Alert data, which is currently in a beta test phase, from pipelines at the \tess\ Science Office and at the \tess\ Science Processing Operations Center. Funding for the \tess\ mission is provided by NASA’s Science Mission directorate. Based on observations made with the ESO 3.6m  telescope at La Silla Observatory under programme ID 072.C-0488, 183.C-0972, and 192.C-0852. This work has made use of data from the European Space Agency (ESA) mission {\it Gaia} (\url{https://www.cosmos.esa.int/gaia}), processed by the {\it Gaia} Data Processing and Analysis Consortium (DPAC, \url{https://www.cosmos.esa.int/web/gaia/dpac/consortium}). Funding for the DPAC has been provided by national institutions, in particular the institutions participating in the {\it Gaia} Multilateral Agreement. We acknowledge the traditional owners of the land on which the AAT stands, the Gamilaraay people, and pay our respects to elders past and present. We are also very grateful to the referee, Steve Howell, for his helpful comments and suggestions, which resulted in an improved paper. We also thank him for his rapid review of the paper.

\end{acknowledgements}

\bibliographystyle{aa} 
\bibliography{DGandolfi_Pi_Men} 

\begin{table*}
\begin{center}
\caption{UCLES RV measurements of \sname.}
\begin{tabular}{lcc}
\hline\hline
\noalign{\smallskip}
$\rm BJD_{TDB}^a$ & RV  & $\pm \sigma$ \\
-2450000          & (\kms)  & (\kms)    \\
\hline
\noalign{\smallskip}
\noalign{\smallskip}
829.993723 &  -0.0410 &   0.0048 \\
1119.251098 &  -0.0674 &   0.0098 \\
1236.033635 &  -0.0792 &   0.0060 \\
1411.325662 &  -0.0858 &   0.0058 \\
1473.267712 &  -0.0800 &   0.0048 \\
1526.081162 &  -0.0930 &   0.0046 \\
1527.082805 &  -0.0898 &   0.0041 \\
1530.128708 &  -0.0879 &   0.0045 \\
1629.912366 &  -0.0927 &   0.0056 \\
1683.842991 &  -0.1005 &   0.0050 \\
1828.188260 &  -0.0674 &   0.0048 \\
1919.099660 &  -0.0350 &   0.0072 \\
1921.139081 &  -0.0373 &   0.0047 \\
1983.919846 &  -0.0028 &   0.0056 \\
2060.840355 &   0.1361 &   0.0048 \\
2092.337359 &   0.2120 &   0.0047 \\
2093.352231 &   0.2094 &   0.0044 \\
2127.328562 &   0.2878 &   0.0059 \\
2128.336410 &   0.2861 &   0.0042 \\
2130.339049 &   0.2899 &   0.0067 \\
2151.292440 &   0.3079 &   0.0052 \\
2154.305009 &   0.3030 &   0.0100 \\
2187.196618 &   0.2857 &   0.0039 \\
2188.236606 &   0.2893 &   0.0037 \\
2189.223031 &   0.2797 &   0.0033 \\
2190.145881 &   0.2835 &   0.0037 \\
2387.871387 &   0.1009 &   0.0036 \\
2389.852023 &   0.0974 &   0.0033 \\
2510.307394 &   0.0417 &   0.0042 \\
2592.126975 &   0.0202 &   0.0032 \\
2599.155380 &   0.0210 &   0.0120 \\
2654.099326 &   0.0188 &   0.0047 \\
2751.918480 &  -0.0117 &   0.0042 \\
2944.224628 &  -0.0434 &   0.0038 \\
3004.075458 &  -0.0321 &   0.0044 \\
3042.078745 &  -0.0440 &   0.0042 \\
3043.018085 &  -0.0463 &   0.0045 \\
3047.050110 &  -0.0408 &   0.0043 \\
3048.097508 &  -0.0444 &   0.0036 \\
3245.311649 &  -0.0697 &   0.0050 \\
3402.035747 &  -0.0669 &   0.0018 \\
3669.244092 &  -0.0863 &   0.0019 \\
\noalign{\smallskip}
\hline
\end{tabular}
\label{Table:AAT}
\\
\end{center}
Notes:\\
$^a$ Barycentric Julian dates are given in barycentric dynamical time. \\
\end{table*}

\begin{table*}
\begin{center}
\caption{\label{Table:HARPS1} HARPS RV measurements of \sname\ acquired with the old fiber bundle. The entire RV data set is available in a machine-readable table in the online journal.}
\begin{tabular}{lcccccc}
\hline\hline
\noalign{\smallskip}
$\rm BJD_{TDB}^a$ & RV  & $\pm \sigma$ &    BIS  & FWHM   & T$_\mathrm{exp}$ & S/N $^b$\\
-2450000          & (\kms)  & (\kms)       &  (\kms) & (\kms)  &      (s)         &   \\
\hline
\noalign{\smallskip}
\noalign{\smallskip}
 3001.830364 & 10.6600 & 0.0014 & -0.0019 & 7.6406 &  109 &  69.0 \\
 3034.607261 & 10.6665 & 0.0008 &  0.0040 & 7.6368 &  200 & 120.7 \\
 3289.869718 & 10.6448 & 0.0012 & -0.0013 & 7.6378 &   60 &  79.6 \\
 3289.870782 & 10.6428 & 0.0011 & -0.0012 & 7.6406 &   60 &  89.4 \\
 3289.871836 & 10.6446 & 0.0011 & -0.0007 & 7.6394 &   60 &  90.7 \\
 3289.872866 & 10.6449 & 0.0011 & -0.0026 & 7.6439 &   60 &  86.5 \\
 $\cdots$      & $\cdots$  & $\cdots$  & $\cdots$ & $\cdots$ & $\cdots$ & $\cdots$ \\
\noalign{\smallskip}
\hline
\end{tabular}
\\
\end{center}
Notes:\\
$^a$ Barycentric Julian dates are given in barycentric dynamical time. \\
$^b$ S/N per pixel at 550 nm. \\
\end{table*}

\begin{table*}
\begin{center}
\caption{\label{Table:HARPS2} HARPS RV measurements of \sname\ acquired with the new fiber bundle.}
\begin{tabular}{lcccccc}
\hline\hline
\noalign{\smallskip}
$\rm BJD_{TDB}^a$ & RV  & $\pm \sigma$ &    BIS  & FWHM   & T$_\mathrm{exp}$ & S/N $^b$\\
-2450000          & (\kms)  & (\kms)       &  (\kms) & (\kms)  &      (s)         &   \\
\hline
\noalign{\smallskip}
\noalign{\smallskip}
 7298.853243 & 10.6750 & 0.0005 &  0.0081 & 7.6856 &  450 & 187.3 \\
 7298.858243 & 10.6747 & 0.0004 &  0.0083 & 7.6842 &  450 & 242.8 \\
 7327.755817 & 10.6744 & 0.0003 &  0.0089 & 7.6870 &  900 & 305.7 \\
 7354.783687 & 10.6674 & 0.0002 &  0.0104 & 7.6867 &  900 & 538.0 \\
 7357.725912 & 10.6727 & 0.0002 &  0.0105 & 7.6858 &  900 & 542.9 \\
 7372.705131 & 10.6664 & 0.0004 &  0.0094 & 7.6825 &  300 & 273.0 \\
 7372.708997 & 10.6662 & 0.0004 &  0.0118 & 7.6822 &  300 & 247.3 \\
 7372.712758 & 10.6654 & 0.0003 &  0.0104 & 7.6831 &  300 & 320.9 \\
 7423.591772 & 10.6630 & 0.0005 &  0.0113 & 7.6796 &  450 & 217.1 \\
 7423.597918 & 10.6628 & 0.0005 &  0.0115 & 7.6782 &  450 & 214.1 \\
 7424.586637 & 10.6645 & 0.0004 &  0.0104 & 7.6814 &  450 & 299.7 \\
 7424.592367 & 10.6643 & 0.0004 &  0.0113 & 7.6816 &  450 & 288.5 \\
 7462.517924 & 10.6612 & 0.0003 &  0.0116 & 7.6822 &  450 & 326.8 \\
 7462.523491 & 10.6612 & 0.0003 &  0.0106 & 7.6816 &  450 & 337.8 \\
 7464.499915 & 10.6616 & 0.0005 &  0.0083 & 7.6812 &  300 & 217.7 \\
 7464.503781 & 10.6627 & 0.0004 &  0.0112 & 7.6818 &  300 & 276.2 \\
 7464.507474 & 10.6611 & 0.0004 &  0.0108 & 7.6820 &  300 & 286.4 \\
\noalign{\smallskip}
\hline
\end{tabular}
\\
\end{center}
Notes:\\
$^a$ Barycentric Julian dates are given in barycentric dynamical time. \\
$^b$ S/N per pixel at 550 nm. \\
\end{table*}

\begin{table*}[!t]
\begin{center}
  \caption{\sname\ system parameters. \label{tab:parameters}}  
  \begin{tabular}{lcc}
  \hline\hline
  \noalign{\smallskip}
  Parameter & Prior$^{(\mathrm{a})}$ & Derived value \\
  \noalign{\smallskip}
  \hline
    \noalign{\smallskip}
    \multicolumn{3}{l}{\emph{\bf{Stellar parameters}}} \\
    \noalign{\smallskip}
    Star mass $M_{\star}$ ($M_\odot$) &  $\cdots$ & \smass[] \\
    Star radius $R_{\star}$ ($R_\odot$) &  $\cdots$ & \sradius[]  \\
    Effective Temperature $\mathrm{T_{eff}}$ (K) & $\cdots$ & \stemp[] \\
    Surface gravity$^{(\mathrm{b})}$ \logg\ (cgs) & $\cdots$  & 4.36\,$\pm$\,0.02 \\
    Surface gravity$^{(\mathrm{c})}$ \logg\ (cgs) & $\cdots$ & 4.33\,$\pm$\,0.09 \\
    Iron abundance [Fe/H] (dex)                   & $\cdots$ & 0.05\,$\pm$\,0.09 \\
    Projected rotational velocity \vsini\ (\kms)  & $\cdots$ & 3.3\,$\pm$\,0.5  \\
    Age (Gyr)                                     & $\cdots$ & 5.2\,$\pm$\,1.1 \\
    \noalign{\smallskip}
    \hline
    \noalign{\smallskip}
    \multicolumn{3}{l}{\emph{\bf{Model parameters of \sname~b}}} \\
    \noalign{\smallskip}
    Orbital period $P_{\mathrm{orb,\,b}}$ (days) &  $\mathcal{U}[2076.0 , 2106.0 ]$ & \Pc[] \\
    Time of inferior conjunction $T_\mathrm{inf,\,b}$ (BJD$_\mathrm{TDB}-$2\,450\,000) & $\mathcal{U}[6528.0 , 6568.0 ]$ & \Tzeroc[]  \\
    $\sqrt{e_\mathrm{b}} \sin \omega_\mathrm{\star,\,b}$ &  $\mathcal{U}[-1,1]$ & \esinc[] \\
    $\sqrt{e_\mathrm{b}} \cos \omega_\mathrm{\star,\,b}$ &  $\mathcal{U}[-1,1]$ & \ecosc[] \\
    Radial velocity semi-amplitude variation $K_\mathrm{b}$ (\ms) & $\mathcal{U}[0,500]$ & \kc[] \\
    \noalign{\smallskip}
    \hline
    \noalign{\smallskip}
    \multicolumn{3}{l}{\emph{\bf{Model parameters of \sname~c}}} \\
    \noalign{\smallskip}
    Orbital period $P_{\mathrm{orb,\,c}}$ (days) &  $\mathcal{U}[ 6.24 , 6.29]$ & \Pb[] \\
    Transit epoch $T_\mathrm{0,\,c}$ (BJD$_\mathrm{TDB}-$2\,450\,000) & $\mathcal{U}[8325.47 , 8325.52]$ & \Tzerob[]  \\ 
    Scaled semi-major axis $a_\mathrm{c}/R_{\star}$ &  $\mathcal{N}[13.11,0.17]$ & \arb[] \\
    Planet-to-star radius ratio $R_\mathrm{c}/R_{\star}$ & $\mathcal{U}[0,0.1]$ & \rrb[]  \\
    Impact parameter $b_\mathrm{c}$  & $\mathcal{U}[0,1]$  & \bb[] \\
    $\sqrt{e_\mathrm{c}} \sin \omega_\mathrm{\star,\,c}$ &  $\mathcal{F}[0]$  & 0 \\
    $\sqrt{e_\mathrm{c}} \cos \omega_\mathrm{\star,\,c}$ &  $\mathcal{F}[0]$  & 0 \\
    Radial velocity semi-amplitude variation $K_\mathrm{c}$ (\ms) & $\mathcal{U}[0,10]$ & \kb[] \\
    \noalign{\smallskip}
    \hline
    \noalign{\smallskip}
    \multicolumn{3}{l}{\emph{\bf{Additional model parameters}}} \\
    \noalign{\smallskip}
    Parameterized limb-darkening coefficient $q_1$  & $\mathcal{N}[0.36,0.1]$ & \qone \\
    Parameterized limb-darkening coefficient $q_2$  & $\mathcal{N}[0.25,0.1]$ & \qtwo \\
    Systemic velocity $\gamma_{\mathrm{AAT}}$  (km s$^{-1}$) & $\mathcal{U}[ -0.20 , 0.20]$ & \ATT[] \\
    Systemic velocity $\gamma_{\mathrm{HS1}}$  (km s$^{-1}$) & $\mathcal{U}[10.53 , 10.88]$ & \HSone[]  \\
    Systemic velocity $\gamma_{\mathrm{HS2}}$  (km s$^{-1}$) & $\mathcal{U}[10.55 , 10.90]$ & \HStwo[] \\
   \noalign{\smallskip}
    RV jitter term $\sigma_{\mathrm{AAT}}$  (\ms) & $\mathcal{U}[0,100]$ & \jATT[]  \\
    \noalign{\smallskip}
    RV jitter term $\sigma_{\mathrm{HS1}}$  (\ms) & $\mathcal{U}[0,100]$ & \jHSone[]  \\
    \noalign{\smallskip}
    RV jitter term $\sigma_{\mathrm{HS2}}$  (\ms) & $\mathcal{U}[0,100]$ & \jHStwo[]  \\
    \noalign{\smallskip}
    \hline 
    \noalign{\smallskip}
    \multicolumn{3}{l}{\emph{\bf{Derived parameters of \sname\,b}}} \\
    \noalign{\smallskip}

    Planet minimum mass $M_\mathrm{b} \sin i_\mathrm{b}$ ($M_\mathrm{Jup}$) & $\cdots$ & \mpc[]  \\
    Semi-major axis of the planetary orbit $a_\mathrm{b}$ (AU) & $\cdots$ & \ac[]  \\
    Orbit eccentricity $e_\mathrm{b}$ & $\cdots$ & \ec[]  \\
    Argument of periastron of stellar orbit $\omega_\mathrm{\star,\,b}$ (deg) & $\cdots$ & \wc[] \\
    Time of periastron passage $T_\mathrm{per,\,b}$ (BJD$_\mathrm{TDB}-$2\,450\,000) & $\cdots$ & \Tperic[]  \\
    \noalign{\smallskip}
    \hline
    \noalign{\smallskip}
    \multicolumn{3}{l}{\emph{\bf{Derived parameters of \sname\,c}}} \\
    \noalign{\smallskip}

    Planet mass $M_\mathrm{c}$ ($M_{\oplus}$) & $\cdots$ & \mpb[]  \\
    Planet radius $R_\mathrm{c}$ ($R_{\oplus}$) & $\cdots$ & \rpb[] \\
    Planet mean density $\rho_\mathrm{c}$ ($\mathrm{g\,cm^{-3}}$) & $\cdots$ & \denpb[] \\
    Semi-major axis of the planetary orbit $a_\mathrm{c}$ (AU) & $\cdots$ & \ab[]  \\
    Orbit eccentricity $e_\mathrm{c}$ & $\cdots$ & 0 (fixed) \\
    Orbit inclination $i_\mathrm{c}$ (deg) & $\cdots$ & \ib[] \\
    Equilibrium temperature$^{(\mathrm{d})}$  $T_\mathrm{eq,\,c}$ (K)  & $\cdots$ &  \Teqb[] \\
    \noalign{\smallskip}
    Transit duration $\tau_\mathrm{14,\,c}$ (hours) & $\cdots$ & \ttotb[] \\
    \noalign{\smallskip}
  \hline
  \end{tabular}
  \begin{tablenotes}\footnotesize
  \item \emph{Note} -- 
                       $^{(\mathrm{a})}$ $\mathcal{U}[a,b]$ refers to uniform priors between $a$ and $b$, and $\mathcal{F}[a]$ to a fixed $a$ value.
                       $^{(\mathrm{b})}$ From spectroscopy and isochrones.
                       $^{(\mathrm{c})}$ From spectroscopy.
                       $^{(\mathrm{d})}$ Assuming zero albedo and uniform redistribution of heat.
\end{tablenotes}
\end{center}
\end{table*}

\end{document}